\documentstyle[prl,aps,twocolumn]{revtex}
\begin{document}
\twocolumn[\hsize\textwidth\columnwidth\hsize\csname @twocolumnfalse\endcsname
\draft
\title{Is Quantum Bit Commitment Really Possible?}
\author{Hoi-Kwong Lo\footnotemark ~and H. F. Chau\footnotemark}
\address{
 School of Natural Sciences, Institute for Advanced Study, Olden Lane,
 Princeton, NJ 08540
}
\date{\today}
\preprint{IASSNS-HEP-96-21; quant-ph/9603004}
\maketitle
\begin{abstract}
 We show that all proposed quantum bit commitment schemes are insecure because
the sender, Alice, can almost always cheat successfully by using an
Einstein-Podolsky-Rosen type of
attack and delaying her measurement until she opens her commitment.
\end{abstract}
\pacs{PACS Numbers: 89.70.+c, 03.65.Bz, 89.80.+h}
]
\narrowtext
\footnotetext[1]{Present Address: BRIMS,
Hewlett-Packard Labs, Filton Road, Stoke Gifford,
Bristol BS12 6QZ, UK. e-mail: hkl@hplb.hpl.hp.com}
\footnotetext[2]{Present Address: Department of Physics,
University of Hong Kong, Pokfulam Road,
Hong Kong. e-mail: hfchau@hkusua.hku.hk}
Work on quantum cryptography was started by S. J. Wiesner in a paper written in
about 1970, but remained unpublished until 1983 \cite{Wiesner}.
Recently, there have been lots of renewed activities in the subject.
The most well-known application of quantum cryptography is the so-called
quantum key distribution (QKD) \cite{bit,Ekert,Today},
which is useful for making communications between
two users totally unintelligible to an eavesdropper. QKD takes advantage of the
uncertainty principle of quantum mechanics: Measuring
a quantum system in general disturbs it.
Therefore, eavesdropping on a quantum communication channel will generally
leave unavoidable disturbance in the transmitted signal which can be
detected by the legitimate users.
Besides QKD, other quantum cryptographic protocols \cite{Rev} have also been
proposed.
In particular, it is generally believed \cite{Today} that quantum mechanics
can protect private information while it is being used for public decision.
Suppose Alice has a secret $x$ and Bob a secret $y$. In a ``two-party
secure computation'' (TPSC), Alice and Bob compute a prescribed function
$f(x,y)$ in such a way that nothing about each party's input is disclosed to
the other, except for what follows logically from one's private input and
the function's output. An example of the TPSC is the millionaires' problem:
Two persons would like to know who is richer, but neither wishes the other to
know the exact amount of money he/she has.

In classical cryptography, TPSC can be achieved either through trusted
intermediaries or by invoking some unproven computational assumptions such as
the hardness of factoring large integers. The great expectation
is that quantum cryptography can get rid of
those requirements
and achieve the same goal using the laws of physics alone.
At the heart of such optimism has been the widespread belief that
{\it unconditionally} secure quantum bit commitment (QBC) schemes
exist\cite{Note}. 
Here we put such optimism into very serious doubt by showing that
{\em all} proposed QBC schemes are insecure: A dishonest party can exploit the
non-local Einstein-Podolsky-Rosen (EPR)\cite{EPR} type
correlations in quantum mechanics to cheat
successfully. To do so, she generally needs to maintain
the coherence of her share of a quantum system by using
a quantum computer. We remark that
all proposed QBC schemes contain an
invalid implicit assumption that some measurements are
performed by the two participants. This is why this EPR-type of
attack was missed in earlier analysis.

\par
Let us first introduce bit commitment. A bit commitment scheme generally
involves two parties, a sender, Alice and a receiver, Bob. Suppose
that Alice has a bit ($b = 0$ or $1$) in mind, to which she would like to
be committed towards Bob. That is, she wishes to provide Bob
with a piece of evidence that she has already chosen the bit and that she
cannot change it. Meanwhile, Bob should not be able to tell from that
evidence what $b$ is. At a later time, however, it must be possible for Alice
to {\it open} the commitment. In other words, Alice must be able to show
Bob which bit she has committed to and convince him that this is indeed
the genuine bit that she had in mind when she committed.

A concrete example of an implementation of bit commitment is
for Alice to write down her bit in a piece of paper, which
is then put in a locked box and handed over to Bob.
While Alice cannot change the value of the bit that she has written
down, without the
key to the box Bob cannot learn it himself. At a later time,
Alice gives the key to Bob, who opens the box and recovers
the value of the committed bit. This illustrative example of
implementation is, however,
inconvenient and insecure. A locked box may be very heavy
and Bob may still try to open it by brute force (e.g. with a hammer).

What do we mean by cheating? As an example, a cheating Alice
may choose
a particular value of $b$ during the commitment phase and
tell Bob {\em another} value during the opening
phase. A bit commitment scheme is secure against
a cheating Alice only if such a fake commitment can be discovered by Bob.
For concreteness,
it is instructive to consider a simple QBC protocol due to
Bennett and Brassard \cite{bit}.
Its procedure
goes as follows:
Alice and Bob first agree on
a security parameter, a positive integer $s$.
The sender, Alice, chooses the value of the committed bit,
$b$. If $b=0$, she prepares and sends Bob a sequence of $s$ photons
each of which is randomly chosen to be
either horizontally or vertically polarized.
Of course, the value of $b$ is kept secret during the commitment phase.
Moreover, the actual polarization of each photon chosen by Alice
is {\it not} announced to Bob.
Similarly, if $b=1$, she prepares
and sends Bob a sequence of $s$ photons each of which is randomly
chosen to be
either 45-degree or 135-degree polarized but once again
the actual polarization
of each photon
is kept secret by Alice.
Bob chooses randomly
between the rectilinear (horizontal and vertical) and diagonal
(45-degree or 135-degree) bases to
measure the polarization of each photon.
This completes the commitment phase.
A simple calculation shows that,
the two density matrices
describing the $s$ photons corresponding to $b=0$ and $b=1$
respectively are exactly the same (and are proportional to
the identity matrix). Consequently, Bob
cannot learn anything about the value of $b$.

At a later time, Alice may {\em open} her commitment by
announcing the value of $b$ and the actual polarization of each of
the $s$ photons. Since Bob has chosen his basis (rectilinear
or diagonal) of measurement randomly for
each photon in the commitment phase, on average, only {\it half} of the
$s$ photons, have
been measured by him in the correct basis.
For those photons, Bob can verify that
Alice's announced polarizations match his measurement results.
Baring EPR attacks, a cheating Alice may, for example,
send rectilinear photons
in the commitment phase (hence commits to $b=0$)
but tell Bob that they are diagonal photons
in the opening phase (hence announces $b=1$).
This is cheating. 
Alice then has to make random
guess for the polarizations of the photons that Bob
has measured along the diagonal basis. Since Bob, on average,
measures $s/2$ photons along the diagonal basis,
Alice, with such a cheating strategy, has only
a probability of $\left( 1/2 \right)^{s/2}$ for success.
See~\cite{BC91} for details.

A key weakness of Bennett and Brassard's
scheme is that
Alice can always cheat successfully by using EPR-pairs.
%%%%%
% 4  begin
%%%%%
%[Recall that an EPR pair is a singlet state of the form
%$ { 1 \over \sqrt{2} } ( | VH \rangle - | VH \rangle )$
%where $| V \rangle $ ($ | H \rangle $ respectively)
%denotes a vertically (horizontally respectively) polarized photon.
%It can alternatively be expressed in the form
%$ { 1 \over \sqrt{2} } ( | +- \rangle - | -+ \rangle ) $
%where $| + \rangle $ ($ | - \rangle $ respectively)
%denotes a 45 $\deg $ (135 $\deg $) polarized photon.
%An EPR pair of photons has the property that if
%the polarizations of both members are measured along
%a common basis (say rectilinear or diagonal), they
%will come out perpendicular to each other.]
%%%%%
% 4  end
%%%%%
Alice can prepare $s$ EPR-pairs of photons
and send a member of each pair
to Bob during the commitment phase.
She
skips her measurements and decides on the value of
$b$ only at the beginning of the opening phase. If she chooses
the value of $b$ to be $0$, she measures the polarization of
the photons in her share along the rectilinear basis.
It is a standard property (the EPR paradox) of an EPR pair that
Alice's measurement result on a photon will always be perpendicular
to Bob's result on the other photon of the pair.
Alice can, therefore, proudly announce those polarizations. Similarly, for
$b=1$, she simply measures along the diagonal basis
and proceeds in a similar manner.
There is no way for Bob to detect this attack.

Bennett and Brassard noted this weakness in the same paper in
which they proposed their scheme\cite{bit}.
Nonetheless, new QBC schemes have been proposed and
it has been generally accepted in the literature
\cite{Today,BC91,BCJL} that they defeat an EPR-type of attack.
Our goal here is to demonstrate that, contrary to
popular belief,
precisely the same type of EPR attack defeats all proposed QBC schemes.

All proposed schemes
involve only one-way communications from Alice to Bob.
On the conceptual level, they all involve Alice sending two
quantum systems to Bob, one during the commit phase
and the other during the opening
phase. [%%%%%
%%%%%%%%% 2  begin
%%%%%%%%%%%%%
%Actually, in some of the proposed
%schemes, part of the communications between Alice and Bob
%is done through a classical channel.
%%%%%
% 2  end
%%%%%
There is no loss of generality in our analysis in considering
quantum communications alone since classical communications
is just a special case of quantum communications.]
More precisely,
the general procedure of
any proposed QBC scheme
can be rephrased in the following manner:

(1) Alice chooses the value of a bit $b$ to which she would
like to be committed
towards Bob. If $b =0$, she prepares a state
\begin{equation}
 |0 \rangle = \sum_i \alpha_i | e_i \rangle_A \otimes | \phi_i \rangle_B ,
\label{zero}
\end{equation}
where $ \langle e_i  | e_j \rangle_A = \delta_{ij}$ but
the normalized states $| \phi_i \rangle_B $'s
are not necessarily orthogonal to each other.
Similarly, if $b=1$, she prepares a state 
\begin{equation}
 |1 \rangle = \sum_j \beta_j | e'_j \rangle_A \otimes | \phi'_j \rangle_B ,
\label{one}
\end{equation}
where $ \langle  e'_i | e'_j \rangle_A = \delta_{ij}$
but $| \phi'_j \rangle_B$'s are not necessarily orthogonal to each other.

Both Alice and Bob are supposed to know the states
$ |0 \rangle$ and $|1 \rangle$. This implies, in particular, that
both of them know the states
$|\phi_i \rangle_B$ and
$|\phi'_j \rangle_B$.

(2) An {\it honest} Alice is now supposed to make a measurement on the first
register and determine the value of $i$ if $b=0$ ($j$ if $b=1$).

(3) Alice sends the second register to Bob as a piece of evidence
for her commitment.

(4) At a later time, Alice opens the commitment by declaring the value
of $b$ and of $i$ or $j$.

(5) Bob
performs measurements on the second register
to verify that Alice has indeed committed to the genuine bit.
More precisely, the data received from Alice (the values of $b$
and also
$i$ or $j$) should be correlated with Bob's experimental
results on the second register. If such expected correlations do
appear, Bob accepts that Alice has executed the protocol
honestly. Otherwise, Bob suspects that Alice is cheating.

We emphasize that all proposed QBC schemes follow the five-step
procedure described above.
For instance, Bennett and Brassard's scheme described earlier
falls into this class if
we give Bob the liberty to store
up his photons and measure them
only after the opening (step 4) of the commitment by Alice.
But, if Alice can cheat against even such a powerful Bob,
clearly she can cheat against Bob who has no such storage capability.
%%%%%
% 1 begin
%%%%%
%Therefore, our proof clearly applies to
%Bennett and Brassard's scheme.
%%%%%
% 1 end
%%%%%

Our proof of insecurity of QBC goes as follows:
First of all, in order that Bob cannot tell what
$b$ is, the second register (the quantum system
that Bob receives during the commit phase) must contain very little
information about which bit Alice has committed to.
As a start, let us consider the {\it ideal} case in which
the second register contains absolutely no information about
the value of $b$.
%%%%%
% 3 begin
%%%%%
[Bennett and Brassard's scheme \cite{bit}
and Ardehali's scheme \cite{Ard} are ideal whereas
Brassard and Cr\'{e}peau's scheme \cite{BC91} and the most well-known BCJL
scheme\cite{BCJL} are non-ideal.
We will come to the non-ideal case near the end of
this Letter.]
%%%%%
% 3 end
%%%%%
In the ideal case, to ensure that
Bob has no information about the committed bit $b$, the density matrices
describing the
second register associated
with the bits $0$ and $1$ are the same. i.e.,
\begin{equation}
{\rm Tr}_A |0 \rangle \langle 0| \equiv  \rho^B_0 =
\rho^B_1 \equiv  {\rm Tr}_A |1 \rangle \langle 1 |.
\label{partial}
\end{equation}

It then follows from the Schmidt decomposition\cite{Jozsa} that
\begin{equation}
|0 \rangle  = \sum_k \sqrt{\lambda}_k 
| \hat{e}_k \rangle_A \otimes | \hat{\phi}_k \rangle_B,
\label{polar}
\end{equation}
and
\begin{equation}
|1 \rangle  = \sum_k \sqrt{\lambda}_k 
| \hat{e}'_k \rangle_A \otimes | \hat{\phi}_k \rangle_B,
\label{polar1}
\end{equation}
where $\{| \hat{e}_k \rangle_A \}$, $\{| \hat{e}'_k \rangle_A\}$
and  $ \{| \hat{\phi}_k \rangle_B \}$ 
are orthonormal bases of the corresponding Hilbert spaces and
$\lambda_k $'s are the eigenvalues of the reduced density
operator,
${\rm Tr}_A |0 \rangle \langle 0|= {\rm Tr}_A |1 \rangle \langle 1 |$.
Notice that the $\lambda_k $'s and $| \hat{\phi}_k \rangle_B$'s are
the same for the two states and the only difference lies in
Alice's system $| \hat{e}_k \rangle_A$'s vs $| \hat{e}'_k \rangle_A$'s.
Now consider the unitary transformation
$U_A$ which maps $| \hat{e}_k \rangle_A$ to $| \hat{e}'_k \rangle_A$.
It clearly maps $ |0 \rangle $ to $|1 \rangle $.
Note that the transformation $U_A$ acts on Alice's system {\em alone} and yet
rotates $ |0 \rangle $ to $|1 \rangle $. That is, Alice can apply $U_A$ without
Bob's help. Therefore, Alice can cheat
by changing $b=0$ to $b=1$ in the opening phase.

More concretely, consider the following cheating strategy: In the
first step, Alice always prepares
$|0 \rangle$ corresponding to $b=0$. She then skips the second (measurement)
step and sends the second register to Bob as prescribed in the third
step. She decides on the value of $b$ to announce only
in the beginning of the opening phase (step 4). Should she now choose
$b$ to be zero, she executes the protocol honestly.
On the other hand, if she now chooses
$b$ to be one, she applies the unitary transformation $U_A$ to rotate
$|0 \rangle$ to $|1 \rangle$ and executes the protocol for $b=1$ instead.
Consequently, Alice can always cheat successfully. Notice that
Alice is able to cheat primarily because she can {\it delay} her measurement
until step four.
To do so,
Alice generally needs
a quantum computer. While it
is a challenging technological feat to build a quantum
computer, it is
not forbidden by the laws of quantum physics.
The possibility of a dishonest
Alice skipping the second step (i.e., delaying her measurements) was not
considered in Ref.~\cite{BCJL}.
This was the chief reason why earlier researchers
came to the erroneous conclusion that
the BCJL scheme is provably unbreakable.

In the above discussion, we have assumed the ideal situation
in which Bob has absolutely
no information about the value of $b$ during the commitment phase and
hence the density matrices
describing the
second register
for the two cases $b=0$ and $b=1$ are the same. (See Eq.\ (\ref{partial}).)
However, Brassard and Cr\'{e}peau's scheme\cite{BC91} and
the BCJL scheme\cite{BCJL} are non-ideal in the sense that they
violate  Eq.\ (\ref{partial}) slightly and
give Bob
some probability of distinguishing between $ \rho^B_0$
and $\rho^B_1$. Intuition seems to
indicate that this is not going to change our conclusion:
On the one hand, if Bob has
a large probability of distinguishing between the
two states, the scheme will be unsafe against a cheating Bob.
On the other hand, if Bob has only a very small probability of
distinguishing between
the two states, clearly the two density matrices
$ \rho^B_0$ and $\rho^B_1$ must be close to
each other in some sense and essentially the same physics should apply.

Following Mayers \cite{Mayers}, we now consider the non-ideal case when
$\rho^B_0 \not= \rho^B_1$.
The closeness
between two states of $B$ specified by the two
density matrices $ \rho^B_0$ and $\rho^B_1$, is commonly described
by the concept {\em fidelity} \cite{Jozsa1}
which can be defined in terms of
{\em purifications}. Imagine a system $A$ attached to Bob's system $B$.
There are many pure states $|\psi_0 \rangle $ and $|\psi_1 \rangle $
on the composite system such that
\begin{equation}
{\rm Tr}_A \left( | \psi_0 \rangle \langle \psi_0 | \right) = \rho^B_0 
\mbox{~~~~~~and~~~~~}
{\rm Tr}_A \left( | \psi_1 \rangle \langle \psi_1 | \right) = \rho^B_1.
\label{pure}
\end{equation}
The pure states $| \psi_0 \rangle$ and $| \psi_1 \rangle$ are
called the purifications of the density matrices $\rho^B_0 $ and $\rho^B_1 $.
The fidelity can be defined as
\begin{equation}
F(\rho^B_0, \rho^B_1) = {\rm max} | \langle \psi_0|\psi_1  \rangle  |
\label{fidelity1}
\end{equation}
where the maximization is over all possible purifications.
$0 \leq F \leq 1$. $F=1$ if and only if $\rho^B_0 = \rho^B_1$.
We remark that for any fixed purification of $\rho^B_1$, e.g.
$ |1 \rangle$ in Eq.\ (\ref{one}), there exists
a maximally parallel purification of $\rho^B_0$ which satisfies 
Eq.\ (\ref{fidelity1}).

For non-ideal QBC schemes,
the fact that Bob has a small probability for distinguishing between
$\rho^B_0$ and $\rho^B_1$ means that \cite{Mayers}
\begin{equation}
F(\rho^B_0, \rho^B_1) = 1 - \delta
\label{fidelity2}
\end{equation}
for some small $\delta > 0$.
It then follows from Eqs.\ (\ref{fidelity1}) and (\ref{fidelity2})
that, for the state $ |1 \rangle$ given in Eq.\ (\ref{one}), there exists
a purification $|\psi_0  \rangle $ of $\rho^B_0 $ such that
\begin{equation}
| \langle \psi_0 | 1 \rangle  | =
F(\rho^B_0, \rho^B_1)= 1- \delta .
\label{fidelity3}
\end{equation}

The strategy of a cheating Alice for a non-ideal bit commitment
scheme is the same as before.
She prepares the state $| 0 \rangle$ corresponding to
$b=0$ in the first step, skips the second (measurement) step and
sends the second register to Bob as prescribed in the third step.
She decides on the value of $b$ only in the beginning
of the opening phase (step 4). If she now
chooses $b=0$, she simply follows the rule.
If she chooses $b=1$, she applies a unitary transformation to
the quantum system on her share to
obtain the state $| \psi_0  \rangle $ which satisfies Eq.\ (\ref{fidelity3}).
%%%%%
% 5 begin
%%%%%
Such a unitary transformation exists because, as can be
seen in the Schmidt decomposition\cite{Jozsa}, all purifications
$| \phi \rangle_{AB}$ of a fixed density matrix $ \rho_B$ are related
to one another by unitary transformations acting on $A$ {\em alone}
and $A$ is in Alice's hands.
%%%%%
% 5 end
%%%%%
Notice that if Alice had been honest, she would have prepared
$| 1  \rangle $ in the first step instead. (See Eq.\ (\ref{one}).)
Nonetheless, since $| \psi_0 \rangle $ and $| 1 \rangle$ are so
similar to each other (See Eq.\ (\ref{fidelity3}).), Bob
clearly has a hard time in detecting
the dishonesty of Alice.
Therefore, Alice can cheat successfully with a very large probability.

%%%%%
% 6 begin
%%%%%
%In summary, we put the widespread
%belief that quantum mechanics can protect private information
%during public decision\cite{Today,BCJL,Ard,CS,BBCS,M,Ar,Yao} in very
%serious doubt by
%showing that all proposed schemes\cite{bit,BC91,BCJL,Ard} for QBC
%are insecure. We manage to do so despite well-known claims of ``provably
%unbreakable'' schemes for QBC.
%The reason is that all those schemes implicitly
%assume that some measurements are made by
%the two participants.
%A cheater, Alice, with a quantum computer
%can exploit this loophole and use non-local EPR-type correlations
%in quantum mechanics to cheat.
%In contrast,
%thanks to the quantum no-cloning theorem \cite{Zurek}
%QKD remains secure\cite{Exp,New}.
%%%%%
% 6 end
%%%%%
We thank helpful discussions with
M. Ardehali, C. H. Bennett, G. Brassard,
C. Cr\'{e}peau, D. P. DiVincenzo, L. Goldenberg,
R. Jozsa, J. Kilian, D. Mayers,
J. Preskill, P. Shor, T. Toffoli and F. Wilczek
after the completion of an earlier version of this Letter.
This work is supported in part by DOE grant DE-FG02-90ER40542.

{\it Notes added}: 
The insecurity of the BCJL scheme \cite{BCJL} has
also been investigated independently by Mayers \cite{Mayers}.
More recently, Mayers\cite{Mayers2} has generalized the above
result to prove that all quantum bit commitment schemes,
including ones that involve two-way (quantum) communications
between Alice and Bob, are insecure.
The same result and the impossibility of {\it ideal} quantum
coin tossing are discussed in our recent preprint\cite{LoChau2}.
The impossibility of some other quantum protocols has recently
been demonstrated by Lo\cite{Lo}.
These surprising discoveries constitute a major
setback to quantum cryptography. The exact boundary to the
power of quantum cryptography remains an important subject for
future investigations.


\begin{references}
\bibitem{Note} Various quantum bit commitment
schemes \cite{bit,BC91,BCJL,Ard} have been proposed
and at least one of them, the so-called BCJL scheme,
is even claimed to be provably unbreakable \cite{BCJL}.
Quantum bit commitment is an important protocol from which
one can construct quantum coin tossing \cite{bit,BC91,BCJL,Ard}, quantum
oblivious mutual identification \cite{CS} and
quantum oblivious transfer \cite{BBCS,Cr94,MS,M,Ar,Yao}.
Kilian \cite{Kilian} has shown that,
in classical cryptography, oblivious transfer forms the basis
of many other protocols including two-party secure computations
\cite{Kilian}.
This chain of arguments
seems to suggest that
quantum bit commitment alone
is sufficient for implementing two-party secure computations,
thus solving a long standing problem in cryptography.
\bibitem{Mayers2} D. Mayers, ``Unconditionally secure quantum bit commitment
is impossible,'' accepted for publication in PRL.
\end{references}
\end{document}